\documentstyle[aps,eqsecnum]{revtex}

\def\m@thcombine#1#2{%
  \setbox0=\hbox{$#1$}
  \setbox1=\hbox{$#2$} 
  \ifdim\wd0>\wd1
    \setbox0=\hbox to\wd1{\hss\box0\hss}
  \else
    \setbox1=\hbox to\wd0{\hss\box1\hss}
  \fi
  \mathop{\vcenter{
    \offinterlineskip\box0\box1}}}
\def\lesim{\m@thcombine<\sim}
\def\gesim{\m@thcombine>\sim}
\begin{document}

%%%%%%%%%%%%%%%%%%%%%%%%%%%%%%%%%%%%%%%%%%%%%%%%%%%%%%%%%%
\draft
\title{ DYNAMICAL STRUCTURE OF THE CHIRAL QCD VACUUM \\                        
        IN THE ZERO MODES ENHANCEMENT QUANTUM MODEL }

\author{V. Gogohia$^{a,b}$, H. Toki$^a$, T. Sakai$^a$ and Gy. Kluge$^b$  }

\address{$^a$Research Center for Nuclear Physics (RCNP), Osaka University \\
          Mihogaoka 10-1, Ibaraki, Osaka 567-0047, Japan  \\
         $^b$HAS, CRIP, RMKI, Depart. Theor. Phys., P.O.B. 49, 
    H-1525 Budapest 114, Hungary \\ email address: gogohia@rcnp.osaka-u.ac.jp and gogohia@rmki.kfki.hu } 

\maketitle

\begin{abstract}
Using the effective potential approach for composite operators we have formulated the quantum model of the QCD vacuum. It is
based on the existence and importance of the nonperturbative $q^{-4}$-type dynamical, topologically nontrivial excitations of the gluon field configurations.
The QCD vacuum is found stable since the vacuum energy density has no imaginary
part. Moreover, a possible ground (stationary) state of the nonperturbative Yang-Mills (quenched QCD) vacuum is discovered. The vacuum energy density at
stationary state depends on a scale at which nonperturbative effects become
important. The quark part of the vacuum energy density depends
in addition on the constant of integration of the corresponding  Schwinger-Dyson equation. The value of the above mentioned scale is determined from the bounds for the pion decay constant in the chiral limit. 
Our value for the chiral QCD vacuum energy density is one
order of magnitude bigger than the instanton based models can provide while a
fair agreement with recent phenomenological and lattice results for the chiral 
condensate is obtained.                                                        
\end{abstract}


%\vfill

%\eject

\section{Introduction }

The nonperturbative QCD vacuum has a very rich dynamical
and topological structure. It is a very complicated medium and its dynamical and topological complexity means that its structure can be organized at various levels (classical, quantum) and it can contain many different components and ingredients which contribute to the vacuum energy density, the one of main characteristics of the QCD ground state.  
Many models of the QCD vacuum involve some extra classical color field 
configurations such as randomly oriented domains of constant color magnetic 
fields [1], background gauge fields, averaged over spin and color [2], stochastic colored background fields [3], etc (see also Ref. [4] and references therein). The most elaborate random and interacting instanton liquid models (RILM and 
IILM) of the QCD vacuum [5] is based on the existence of the topologically nontrivial instanton-type fluctuations of gluon fields, which are solutions to the 
classical equations of motion in Euclidean space [6-8]. 

Today there are no doubts left that the dynamical mechanisms
of the important nonperturbative quantum phenomena such as quark confinement and dynamical (or equivalently spontaneous) chiral symmetry breaking (DCSB) are closely related to the complicated topologically nontrivial structure of 
the QCD vacuum [9-12]. On the other hand, it also becomes clear that the nonperturbative infrared (IR) dynamical singularities, closely related to the nontrivial vacuum structure, play an important role in the large distance behaviour 
of QCD [13,14]. For this reason, any correct nonperturbative model of quark confinement and DCSB necessarily turns out to be a model of the true QCD vacuum and the other way around. Our model of
the true QCD ground state is based on the existence and importance of such kind
of the nonperturbative, quantum excitations of the gluon field configurations which effectevely correctly can be described by the $q^{-4}$ behaviour of the full gluon propagator in the IR domain (at small $q^2$). It describes the zero modes enhancement (ZME) dynamical effect in QCD at large distances [15-19] (for additional references see Ref. [19]). These excitations are also
topologically nontrivial in comparison with the free gluon structure, $q^{-2}$.

In this context, we note that the attractive classical model of the QCD vacuum as a condensation
of the color-magnetic monopoles (QCD vacuum is a chromomagnetic superconductor)
proposed by Nambu, Mandelstam and 't Hooft and developed by Nair and Rosenzweig
(see Ref. [20] and references therein) as well as the classical mechanism of   
the confining medium [21] and an effective theory for the QCD vacuum proposed  
in [22], also invoke $q^{-4}$ behaviour of the gluon fields in the IR. 
Let us underline that without $q^{-4}$ component in
the decomposition of the full gluon propagator it is impossible to obtain the  
area law for static quarks (indicative of confinement)
within the Wilson loop approach [23]. This behaviour of the full gluon propagator in the IR is also required to derive the heavy quark potential within the recently proposed exact renormalization group approach [24].    

In the next section for the sake of the reader's convenience we provide a brief
review on the dynamical equations approach to nonperturbative QCD, since our quantum model of the QCD vacuum is its direct consequence.

\section{Dynamical equations approach to nonperturbative QCD }

Our approach to nonperturbative QCD is based on solutions to
the quark and ghost Schwinger-Dyson (SD) quantum equations of motion which should be complemented by the investigation of the corresponding Slavnov-Taylor (ST) identities [19,25,26]. If it is true that QCD is an IR unstable theory    
(i. e. if it has no IR stable fixed point) [27] then the low-frequency modes of
the Yang-Mills (YM) fields might be enhanced due to nonperturbative IR divergences [28]. So the full gluon propagator can diverge faster than the free one at 
small momenta [15-29]      
                                             
\begin{equation}
D_{\mu\nu} (q) \sim (q^2)^{-2}, \qquad q^2 \rightarrow 0,
\end{equation}
which might be equivalently refered to as the strong coupling regime [27].
If indeed the low-frequency components of the virtual fields in the true vacuum
have a larger amplitude than those of the bare (perturbative) vacuum [15], then
the Green function, describining the propagation of a single quark in the true 
QCD vacuum, should be reconstucted on the basis of this effect. It is important
to understand that a possible ZME effect (2.1) is our primery dynamical assumption.\footnote{However, let us remind the reader that after the pioneering papers of Mandelstam in the covariant (Landau) gauge [15] and Baker, Ball and Zachariasen in the axial gauge [16], the consistancy of the singular asymptotics (2.1) with direct solution to the SD equation for the full gluon propagator in the 
IR domain was reapeatedly confirmed (see for example Refs. [17,19,29] and references therein).} We consider this effect as a very similar confining ansatz for the full gluon propagator in order to use it as input information for the quark and ghost SD equations as well as in the corresponding quark-gluon ST identitiy.  

Such a singular behaviour of the full gluon propagator in
the IR region requires the introduction of a small IR regularization parameter $\epsilon$, in order to define the initial equations in the IR by the dimensional regularization method [30] within the distribution theory [31]. This 
yeilds the regularization expansion for the above mentioned strong IR singularity as follows (of course, under integrals in four dimensional Euclidean space) 
[19,25,26,31]

\begin{equation}
(q^2)^{-2 + \epsilon} = {\pi^2 \over \epsilon} \delta^4 (q) + finite \ terms,
\qquad \epsilon \rightarrow 0^+
\end{equation}
and the terms of order $\epsilon$ are not shown here for simplicity. Because of
this, the quark propagator and other Green's functions become dependent, in general, on this IR regulation parameter $\epsilon$, which is to be set to zero at
the end of computations, $\epsilon \rightarrow 0^+$. For the sake of brevity, this dependence is always understood but not indicated explicitly.              

 There are only two different types of
behaviour of the quark propagator with respect to $\epsilon$ in
the $\epsilon \rightarrow 0^+$ limit.
 If the quark propagator does not depend
on the $\epsilon$ - parameter in the $\epsilon \rightarrow 0^+$
limit then one obtains the IR regularized
(from the very beginning) quark propagator. In this
case quark confinement is understood as the disappearance of the
quark propagator pole on the real axis at the point $p^2 = m^2$,
where $m$ is the quark mass. Such an understanding (interpretation) of quark confinement comes, apparently from Preparata's massive quark model (MQM) [32] in 
which external quark legs were approximated by entire functions. A quark propagator may or may not be an entire function, but in any case the pole of the first order
(like the electron Green's function has in QED) disappears (see Ref. [19] and 
references therein). However, $the \ absence \ of \ the \ pole-type \ singularities$ in the quark Green's functions is only the $first \ necessary$ condition 
of the quark confinement at the fundamental (microscopic) quark-gluon level. At
hadron (macroscopic) level there exists the $second \ sufficient$ condition, 
namely the corresponding Bethe-Salpeter (BS) equation for the bound-states should have the $discrete \ spectrum \ only$ [33] in order to prevent quarks to appear in asymptotically free states. Consequently, the meson-quark decay amplitude vanishes in the IR limit $\epsilon \rightarrow 0^+$, consistent with confinement [18].                                                                      
%At nonzero temperatures and densities, for example in quark-gluon plasma (QGP)%, the bound-states will be dissolved, but, nevertheless the first necessary co%ndition still remains valid, of course.                                       
Thus in general case the confinement criterion consists of the two above formulated conditions. This definition generalises the linearly rising potential between heavy quarks 
since it is relevant not only for light quarks but for heavy quarks as well.   
On the other hand, a quark propagator can vanish after the removal ($\epsilon \rightarrow 0^+$) of the IR regularization parameter $\epsilon$. A  vanishing quark propagator is also a direct manifestation of quark confinement (see again Ref. [19] and references therein). Such understanding of quark confinement comes, apparently, from two-dimensional QCD with $N_c$ large limit [33]. 

Here a few remarks are made in advance. From our approach to nonperturbative QCD, it follows that the first nontrivial approximation to 
the quark-gluon vertex (as it is required by the correct treatment (2.2) of the
above mentioned strong IR singularity (2.1) by the distribution theory [31] within the dimensional regularization method [30]) is the quark-gluon vertex at zero momentum transfer. It depends on four independent form factors which should 
be determined within the corresponding ST identity. Obviously, in this case there is no sense to speak about potential at all (in particular linearly rising one between heavy quarks) and it is necessary only to speak about the large scale structure of the QCD vacuum and its revelance to quark confinement and other 
nonperturbative effects in QCD. In other words, nobody can believe that the potential concept is useful in this case.\footnote{One of the authors (V.G.) is 
grateful to G.'t Hooft for the discussion of the properties of the $q^{-4}$ singularities from this point of view.} This vertex might be approximated by point-like vertex only for heavy quarks when one neglects corrections to the vertex 
induced by virtual gluons. In this case the linearly rising potential becomes relevant and that is why it was "seen" in lattice QCD [34]. This should be considered as strong lattice evidence (though not direct) of the existence and importance of $q^{-4}$-type excitations of gluon field configurations in the QCD vacuum. As it has been already mentioned above, without this component in the decomposition of the full gluon propagator in continuum theory it is impossible to 
"see" linearly rising potential by lattice simulations not involving 
some extra (besides gluons and quarks) degrees of freedom.  

 We develop a method for the extraction of the IR finite (regularized) Green's 
functions in QCD. The IR finiteness of the Green's functions means that they exist as $\epsilon \rightarrow 0^+$. For this purpose, we have worked out a renormalization program in order to cancel all IR nonperturbative divergences, which
makes it possible to explicitly show that all Green's functions
are IR multiplicative renormalizable (IRMR). We have also shown that for the covariant gauges the complications due to ghost contributions can be considered in our approach. A closed set of equations for the IR finite from the very beginning quark propagator ($S \equiv \bar S$) and other IR renormalized quantities 
in the quark sector is worked out to be [19,25,26]

\begin{eqnarray}
S^{-1}(p) &=& S_0^{-1}(p) + \bar \mu^2 \Gamma_{\mu} (p, 0)S(p) \gamma_{\mu} + S^{UV}(p), \\
\Gamma_{\mu} (p, 0)  &=& i d_{\mu} S^{-1}(p) - S(p) \Gamma_{\mu} (p, 0)S^{-1}(p)  + \Gamma_{\mu}^{UV}(p, 0)
\end{eqnarray}
with $d_{\mu} = d / dp_{\mu}$. Here $S_0(p)$ and $S(p)$ are the free and full quark propagators, respectively while $\Gamma_{\mu} (p, 0)$ is the corresponding
quark-gluon full vertex at zero momentum transfer. Evidently, the second terms 
in Eqs.(2.3-2.4) are due to the deep IR ($q^{-4}$),
while the third terms are due to the ultraviolet (UV) components in the decomposition of the full gluon propagator. The characteristic IR renormalized mass scale parameter, $\bar \mu^2$, determines in general the physical scale of nonperturbative dynamics within our approach. In particular, it is directly responsible for dynamical breakdown of chiral symmetry at the fundamental quark level.\footnote{Precisely this interpretation will be put into the heart of our scale-setting scheme in order to perform numerical calculations, see section 4 below.} Indeed, if it is zero ($\bar \mu^2=0$) then from Eq. (2.3) it follows that there is no contribution to the quark mass generated function (quark self-energy) 
from the deep IR (confinement) region. In other words, the nonperturbative phase does not exist at all in this case. It comes from the initial mass
scale parameter $ \mu^2$ which is necessary to introduce due to the $q^{-4}$ behaviour of the full gluon propagator in the deep IR region. The IR finite quark
wave function renormalization constant which should multiply the free
quark propagator in Eq.(2.3) is to be set to one without loosing generality because of the above mentioned IRMR property of our approach [19,25]. 

We first approximate the exact quark SD equation (2.3-2.4) by its deep IR (confinement) piece only assuming that precisely this term is mainly responsible for
nonperturbative effects in
QCD in particular quark confinement. That is why we do not present the explicit, rather complicated expressions for the corresponding UV terms in Eq. (2.3-2.4). Because of the same reason, the quark wave function UVMR constant is also set to one since the confinement piece of the quark SD equation is
free from UV divergences. The nontrivial UVMR program should be done either from the very beginning or after completing our IRMR program when one takes into
account the UV terms as well. Let us note in advance that approximating the full gluon propagator by its deep IR (confinement) piece only in the whole range, 
we nevertheless will obtain a solution for the dynamically generated quark mass
function, which manifests the existence of the effective  
boundary value momentum (see below).                                           
                                                               
Introducing then the appropriate Euclidean dimensionless momentum variable $x= p^2/ \bar \mu^2$, and doing some algebra, the quark SD equation (2.3-2.4) (without UV pieces) in terms of the quark propagator $iS(p) = \hat p A(p^2) - B(p^2)$ form factors, becomes [19,25,26]

\begin{eqnarray}
xA' &=& - (2+x)A - 1 - m_0 B, \\                
2BB' &=& - 3 A^2 - 2(B - m_0A)B,               
\end{eqnarray}
where $A,B \equiv A(x),B(x)$ and the prime denotes the derivative with respect
to $x$ and $\hat p = \gamma_{\mu} p_{\mu}$. Let us emphasize that the obtained 
system of equations (2.5-2.6) certainly leads                                  
to dynamical breakdown of chiral symmetry (quark mass generation) since $ a \  
chiral \ symmetry \ violating$ solution ($m_0=0, \ A(x) \ne 0, \ B(x) \ne 0$) $
is  \ only \ allowed$, while $a \ chiral \ symmetry \ preserving$ solution ($m_0=B(x)=0, \ A(x) \ne 0$) $is \ forbidden$. 

It is easy to check that 
solutions to this system in the chiral limit ($m_0=0$) are 

\begin{eqnarray}
A(x) &=& x^{-2}( 1 - x - e^{-x}), \\                                           
B^2(x_0, x) &=& 3e^{-2x} \int^{x_0}_x dx'e^{2x'} A^2(x'),                      
\end{eqnarray}
where $x_0=p_0^2/ \bar \mu^2$ is the corresponding (arbitrary at this stage) constant of integration. It plays the role of UV cut-off (dimensionless) while $p_0$ is the UV cut-off in momentum space. It is easy to see that obtained solution for the quark propagator is regular at zero, has no pole-type singularities 
(indicative of confinement). This remains valid for the nonchiral case as well.
As it has been already underlined above, it corresponds
to dynamical breakdown of chiral symmetry (quark mass generation). It is also nonperturbative ( it can not be expanded
in powers of the coupling constant) and the function $A(x)$ automatically approaches the free propagator at infinity (asymptotic freedom). In the solution of 
the differential equation for it, the constant of integration was put to infinity from the very beginning unlike for the quark mass 
generated function $B(x)$.                                         

The remarkable feature of
the solution (2.8) for the dynamically generated quark mass function is that it
exhibits an algebraic branch points at $x=x_0$ and at infinity ($x \rightarrow \infty$ at fixed $x_0$), which are caused by the inevitable ghost contributions
in the covarint gauge. That is why the limit $x_0 \rightarrow 0$ is unphysical 
since the dynamically generated quark mass function $B^2(x_0, x)$ becomes pure 
imaginary which contradicts the idea that the confining particle should have no
imaginary part [35]. On the other hand, this limit is achieved when $\bar \mu^2\rightarrow \infty$ (at fixed $p_0$), but this is impossible since the physical scale of the nonperturbative dynamics $\bar \mu$ is either finite (then the 
nonperturbative phase exists) or zero (only the perturbative phase survives, see Eq. (2.3)), so it can not be arbitrarily large. The perturbative (UV) limit  
$x_0 \rightarrow \infty$ in terms of the constant of integration $x_0=p_0^2/ \bar \mu^2$ can be achieved by two different ways. First, it is recovered at 
fixed $p_0^2$ when the renormalized mass scale parameter $\bar \mu^2$ goes to zero ($\bar \mu^2 \rightarrow 0$) indicating that in this limit there is no nonperturbative phase at all and only the perturbative phase remains, see Eq. (2.3). Second, the UV limit is also 
recovered at fixed $\bar \mu^2$, but the UV cut-off $p_0$ goes to infinity ($p_0 \rightarrow \infty$). However, one can not put automatically
$x_0 = \infty$ for the dynamically generated quark mass function $B^2(x_0, x)$
since the integral (2.8) with $x_0 = \infty$ for any finite $x$ in particular $x=0$ does not exist at all. Hence, we have to keep the constant
of integration $x_0$ in (2.8) arbitrary but finite in order to obtain a regular
and nontrivial solution for the deep IR region, $x_0 \ge x$. Thus it plays the 
role of the UV cut-off, separating deep IR (confinement) region, where nonperturbative effects become dominant, from the intermadiate and UV regions where they might be neglected. From the solution (2.8) it also follows that extrapolating it
to infinity ($x \rightarrow \infty$), it should be simultanously accompanied by
the UV limit in terms of the constant of integration  $x_0 \rightarrow \infty$
in order to avoid the influence of the unphysical singularity (the above mentioned branch point at infinity at fixed $x_0$) on our solutions. In other words,
the correct UV limit $x \rightarrow \infty$ requires $x_0 \rightarrow \infty$ and vice versa. In this case the dynamically generated quark mass function (2.8)
identically vanishes in accordance with the vanishing current quark mass in the
chiral limit. In what follows in all integrals containing the quark degrees of freedom (the dynamically generated quark mass function $B^2(x_0, x)$) explicitly, the perturbative (UV) limit shoud be understood in this sense. Thus one 
may conclude, that the analytical properties of our solutions (structure of the
singularities) influence both the IR and UV behaviour of the quark propagator. 
                                                                
 Concluding this section, we note that our solution does not explicitly depend on ghost degrees of freedom as well as on gauge choice. It is worthwile to emphasize in advance, that within our approach [19,25,26] the explicit ghost and gauge dependence is shifted from confinement piece to the UV piece of the full quark propagator (2.3-2.4).

\section{Zero modes enhancement quantum model of the QCD vacuum }

   As it was mentioned in Introduction any correct nonperturbative model of
quark confinement and DCSB necessarily becomes a model of the QCD ground state,
i. e. its nonperturbative vacuum. The effective potential approach for composite operators [36,37] allows us to investigate the QCD vacuum, since in the
absence of external sources the effective potential is nothing
but the vacuum energy density, one of the main characteristics of the nonperturbative vacuum. It gives the vacuum energy density in the form of loops expansion where the number of 2PI vacuum loop (consisting of confining quarks with dynamically generated quark masses and nonperturbative gluons properly regularized
with the help of ghost) is equal to the power of the Plank constant, $\hbar$.  
So here we will establish completely quantum part of the vacuum energy density 
which is due to $q^{-4}$-type nonperturbative, topologically nontrivial excitations of the gluon field configurations in the QCD vacuum.                      
         
It is convenient to start from
the quark part of the vacuum energy density which to leading order (log-loop level $\sim \hbar$)\footnote{Next-to-leading and higher terms (two and more vacuum loops) are suppressed by one order of magnitude in powers of $\hbar$ at least
and are left for consideration elsewhere.}is given by the effective potential for composite operators as follows [36] 

\begin{equation}
V(S) =- i \int {d^np \over {(2\pi)^n}} Tr \left\{
 \ln (S_0^{-1}S) - (S_0^{-1}S) + 1 \right\},
\end{equation}
where $S(p)$ and $S_0(p)$ are the full and free quark propagators,
respectively. Here and everywhere below in this section the traces over space-time and color group indices are understood. Let us note that the effective potential (3.1) is normalized as $V(S_0)= 0$, i. e. the perturbative vacuum 
is normalized to zero. In order to evaluate the effective potential (3.1) we use the well-known expression,

\begin{equation}
 Tr \ln (S_0^{-1}S) = 3 \times \ln det (S_0^{-1}S) =
 3 \times 2 \ln p^2 \left[ p^2 A^2(-p^2) - B^2(-p^2) \right],
\end{equation}
where $p^2 A^2(-p^2) - B^2(-p^2)  = \sqrt {det[-iS(p)]}$.
The factor 3 comes from the trace over quark color
indices. Going over to Euclidean space ($n=4, \ d^4p \rightarrow i d^4p, \quad p^2 \rightarrow -p^2$ ), in terms of dimensionless variables and functions (2.5-2.8), we finally obtain after some algebra ($\epsilon_q = V(A, B)$),

\begin{equation}
\Omega_q = { 1 \over p_0^4} \epsilon_q = {3 \over 8 \pi^2} \times x_0^{-2} I_q(x_0,0),                     
\end{equation}
and

\begin{equation}
I_q(x_0, 0) =
\int \limits_0^{x_0} dx\, x\, \left\{ \ln (x [x A^2(x) +
B^2(x_0, x)])  + 2 x A(x) + 2 \right\},
\end{equation}
where for further aims it is convenient to introduce the quark effective potential at fixed $p_0$, which becomes a function of dimensionless variable $x_0$ only. Here we need to identify the UV cut-off with the constant of integration $x_0$ in order to guarantee that the unphysical singularities (algebraic branch points, mentioned above in section
2) in the $B^2(x_0, x)$ function (2.8) will not affect the effective potential,
which should be always real in order to avoid the vacuum instability [38].
The constant of integration  $x_0$ is related to the scale below which nonperturbative effects become essential. For this reason,  within our 
approach to QCD at large distances in order to obtain numerical values
of any physical quantity, e.g. the pion decay
constant (see below and Ref. [39]), the integration over the whole range
$[ 0, \  \infty )$ reduces to the integration over the nonperturbative region $ \left[ 0, \  x_0 \right]$,  which determines the range of validity of the corresponding solutions (2.7) and (2.8) for the confinement piece of
the full quark propagator. We emphasize that the main contribution to the values of the physical quantities comes from the nonperturbative region
(large distances), whereas the contributions from the
short and intermediate distances (perturbative region), because of less singular behaviour in the IR, can only be treated as perturbative corrections. We
shall confirm this physically reasonable assertion numerically.
          
In order to correctly recover the perturbative (UV) limit, it is necessary to neglect the dynamically generated quark mass function $B^2(x_0, x)$ (2.8) in the
integrand function (3.4). As it was explained above (section 2) the correct perturbative limit $x_0 \rightarrow \infty$ requires the simultaneous UV limit, $x \rightarrow \infty$. Then one obtains that $\Omega_q$ as a function of $x_0$ at upper limit ($x_0 \rightarrow \infty$) converges as
$x_0^{-2} \ln x_0$, satisfying thereby the initial normalization condition of the perturbative vacuum to be zero. At the same time, the quark effective potential (3.3-3.4) as a function of $x_0$ has no stationary state (a local minimum) 
while the regularized gluon part does have (see below). Let us note also that at fixed $\bar \mu$, the corresponding effective potential
$\bar \Omega_q = ( 1 / \bar \mu^4) \epsilon_q$ as a function of $x_0$ has uncorrect UV limit (it diverges as $\ln x_0$ at upper limit, $x_0 \rightarrow \infty$) contradicting thereby the above mentioned normalization condition. Thus $\bar \mu$ can not be fixed. It should be determined from some good physical observable (see section 4 below). The difference in the behaviour of the effective potential (as a function of $x_0$) at fixed $p_0$ or $\bar \mu$ should be traced 
back to its two versions when the variable is the quark propagator, $S$, [36] or the quark self-energy, $\Sigma$, [37], respectively. 

We evaluate the nonperturbative gluon part of the effective potential, which at
the log-loop level is given by [36]

\begin{equation}
V(D) =  { i \over 2} \int {d^np \over {(2\pi)^n}}
 Tr\{ \ln (D_0^{-1}D) - (D_0^{-1}D) + 1 \}, 
\end{equation}

where $D(p)$ is the full gluon propagator and $D_0(p)$ is its  
free (perturbative) counterpart.
The effective potential is normalized as $V(D_0) = 0$, i. e. as in the quark case the perturbative vacuum is normalized to zero.
In a similar way to Eq. (3.2), we obtain

\begin{equation}
 Tr \ln (D_0^{-1}D) = 8 \times \ln det (D_0^{-1}D) =
 8 \times 4 \ln \left[ {3 \over 4 }d(-p^2) + {1 \over 4 } \right],
\end{equation}
where the factor 8 is due to the trace over the gluon colour indices and it becomes zero (in accordance with the above mentioned normalization condition) when
the full gluon form factor is replaced by its free counterpart by setting simply $d(-p^2, a) = 1$ in (3.6). Let us note that this decomposition does not explicitly depend on a gauge choice. Approximating now the full gluon form factor by
its deep IR (confinement) piece,  namely $d(-p^2)= \bar \mu^2 / (- p^2)$
and after doing some algebra in terms of new variables and parameters (2.5-2.8), we finally obtain (in Euclidean space, $n=4$) the following expression for the vacuum energy density due to nonperturbative gluon contributions, 
$\epsilon_g = V(D)$, 
    
\begin{equation}
\epsilon_g = {1 \over \pi^2} q_0^4 z_0^{-2} \times I^a_g(z_0, 0),
\end{equation}
where    

\begin{equation}
I^a_g (z_0, 0) = \int \limits_0^{z_0} dx\, x\, \Bigl\{  \ln x - \ln (x+ 3)     
+ {3 \over 4x} - a  \Bigr\} 
=  {9 \over 2} \ln(1 + {z_0 \over 3})
- {3 \over 4} z_0 - {1 \over 2} z^2_0 \ln(1 + {3 \over z_0})                 
-  {a \over 2} z^2_0
\end{equation}
and $a = (3/4) - 2 \ln 2 = - 0.6363$. In this equation $z_0$ is, of course, the
corresponding UV cut-off which in general differs (i. e. independent) from the 
quark cut-off $x_0$, i. e. $z_0 = q_0^2 / \bar \mu^2$.

 The effective potential at the log-loop level for the ghost degrees of freedom
is [36]

\begin{equation}
V(G) = - i \int {d^np \over {(2\pi)^n}}
Tr\{ \ln (G_0^{-1} G) - (G_0^{-1} G) + 1 \}, 
\end{equation}
where $ G(p)$ is the full ghost propagator and $G_0(p)$ is its  
free (perturbative) counterparts. 
The effective potential $V(G)$ is normalized as 
$V(G_0) = 0$, i.e. here like in quark and gluon cases the energy of the perturbative vacuum is set to zero. Evaluating the ghost term $\epsilon_{gh} = V(G)$ in (3.9) in a very similar way, we obtain
$\epsilon_{gh} = \pi^{-2} k_0^4 y_0^{-2} \times I_{gh}(y_0, 0)$,
where the integral $I_{gh}(y_0, 0)$ depends on the IR renormalized ghost self-energy, which remains arbitrary (unknown) within our approach. We have introduced the ghost UV cut-off $y_0$ as well, $y_0 = k_0^2 / \bar \mu^2$.      

 In principle, we must sum up all contributions in order to obtain total vacuum
energy density. However, $\epsilon_g$ and $\epsilon_{gh}$ are divergent and therefore they depend completely on arbitrary UV cut-offs $z_0$ and $y_0$ which are in general different from $x_0$. We know how to calculate $x_0$ by relating it to some good physical observable (see section 4 below). At this stage we have
no idea how to calculate $z_0$ (see however below). That is why the only way to
calculate $z_0$ is to identify it with $x_0$ by using the arbitrariness of ghost degrees of freedom. Thus the sum $\epsilon_g + \epsilon_{gh}$ should be regularized in order to define finite  $\epsilon^{reg}_g$, which will depend 
only on $x_0$ and at the same time it should be a vanishing function of $x_0$ at $x_0 \rightarrow \infty$ because of the normalization of the perturbative vacuum to zero. Approximating the full gluon propagator by its deep IR (confinement) asymptotics only in (3.6), the nonzero constant $a$ appears in (3.8) which precisely violates the above mentioned normalization condition of the perturbative vacuum to be zero. So it should be additionally subtracted. For this purpose, we decompose the integral (3.8) into the three parts as follows $I^a_g(z_0, 0) =I^a_g(x_0, 0) + I^a_g(z_0, x_0)= I_g(x_0, 0) - (a/2) x_0^2 + I^a_g(z_0, x_0)$. Using the arbitrariness of the ghost term (3.9), let us 
subtract the unknown integral $- (a/2) x_0^2 + I^a_g(z_0, x_0)$ from the gluon
part of the vacuum energy density by imposing  the following condition $\Delta =-(a/2) x_0^2 + I^a_g (z_0, x_0) + I_{gh}(y_0, 0) = 0$ (having in mind that 
$\bar \mu^4 = p_0^4 x_0^{-2} = q_0^4 z_0^{-2} = k_0^4 y_0^{-2}$).     
In fact, we regularize the gluon contribution to the vacuum energy density  
by subtracting unknown term by means of another unknown (arbitrary) ghost 
contribution, i. e. $ \epsilon_g + \epsilon_{gh} = \epsilon^{reg}_g = \pi^{-2} p_0^4 x_0^{-2} I_g (x_0, 0)$ and in what follows superscript "reg" will be omitted for simplicity. It is easy to show that the
above mentioned condition of cancellation $\Delta =0$ of the UV divergences (due to arbitrary $z_0, \ y_0$) is consistent with this definition of the regularized vacuum energy density of the nonperturbative gluon part. At the same time, 
it becomes negative and has a local minimum (see below), while the unregularized contribution $\pi^{-2} p_0^4 x_0^{-2} I^a_g (x_0, 0)$  is always positive (because of constant $a$). Thus our regularization procedure is in agreement with 
general physical interpretation of ghosts to cancel the effects of the unphysical degrees of freedom of the gauge bosons [27,40].
    
The regularized vacuum energy density due to the nonperturbative gluon contributions becomes

\begin{equation}
\Omega_g = { 1 \over  p_0^4} \epsilon_g = {1 \over 2 \pi^2} \times x_0^{-2}  
\Bigl\{ 9 \ln(1 + {x_0 \over 3}) - {3 \over 2} x_0
-  x^2_0 \ln(1 + {3 \over x_0}) \Bigr\},
\end{equation}
where, in a similar way as for quarks, we introduce the gluon effective potential at fixed $p_0$, namely $\Omega_g$. Its behaviour as a 
function of $x_0$ is shown in Fig. 1. It has a few remarkable features. First,
a local minimum appears at\footnote{In our previous publication, Ref. [26],  
the existence of the stationary state in pure gluodynamics (quenched QCD) was 
not, unfortunately, noticed. Though the quark part of the vacuum energy density
was estimated correctly, the nonperturbative gluon contribution was, for the above mentioned reason, substantially underestimated. Stationarity of the nonperturbative YM vacuum was first observed and used for preliminary numerical calculations in Ref. [41].}     
                             
\begin{equation}
x^{min}_0 = 4 \ln(1 + {x^{min}_0 \over 3}) = 2.20
\end{equation}
(stationary condition) 
and it has zero at $x_0^0 = 0.725$. Second it asymptotically vanishes as $x_0 \rightarrow \infty$ in agreement with the normalization condition.\footnote{Simiral to quark case, the gluon effective potential $\bar \Omega_g = ( 1 / \bar \mu^4) \epsilon_g$ at fixed $\bar \mu$, has unccorect UV behaviour (it diverges as $\sim -x_0$ as $x_0 \rightarrow \infty$).}                               
As it was discussed above (section 2) the opposite limit $x_0 \rightarrow 0$ ($\bar \mu^2 \rightarrow \infty$) is unphysical since the nonperturbatibe IR renormalized scale $\bar \mu$ is zero or finite, it can not be 
arbitrarily large. That is why the vacuum energy density in this limit
becomes positive (see Fig. 1). This obviously means that the physical region for parameter $x_0$ is bounded from below, namely $x_0 \ge x_0^0 = 0.725$ in complete agreement with the discussion in section 2. In this region the effective potential due to nonperturbative gluons is always negative. Thus it has no imaginary part (our vacuum is always stable) and at stationary state it
is $\Omega_g (x_0^{min}) = - 0.0263$ (see Fig.1).  Let us also undeline that without the existence of stationary state in the YM vacuum it would be impossible
to calculate this number. It becomes clear now
that this minimum would not be changed if the effective potential $\Omega_g = (1/ q_0^4) \epsilon_g$ would remain a function of $z_0$ in (3.7). What is important indeed, is to subtract the constant $a$ from (3.8) in order to proceed to  
(3.10) as a function of $z_0$ instead of  $x_0$. 
 
  The vacuum energy density for quenced QCD due to nonperturbative gluons at stationary state becomes  

\begin{equation}                                                       
\epsilon_g = \epsilon_g (x_0^{min}, p_0) = - 0.0263 p_0^4.
\end{equation}                   
At first sight it is badly divergent since $p_0$ is formally the UV cut-off in momentum space. However, within our model it is effectively determines the range of validity of the corresponding $q^{-4}$ behaviour of the full gluon propagator in the deep IR region ($p_0 \equiv q_0$) which, obviously can not be arbitrary large. The finiteness of this scale means that it might be interpreted as the effective (possibly confinement) scale of nonperturbative dynamics within our approach. If QCD confines such an effective scale should certainly exist. It is clear that it can not be numerically determined within the YM (quenced QCD) theory alone. However, it seems to us that $ 1 \ GeV$ is a realistic upper bound for the effective scale responsible for nonperturbative dynamics in quenced QCD. Let us underline, that its value has nothing to do with the values chosen to analyze numerical results in phenomenology or lattice approach (for example, 
$1 \ GeV$ or $2 \ GeV$) which have no physical sense and are simply convention,
while our scale has a direct and clear physical meaning as separating in general the nonperturbative phase from the perturbative one.                         

The total vacuum energy density to leading order (as it is determined by ZME quantum model of the QCD vacuum) finally becomes  

\begin{equation}                                                               
\epsilon = \epsilon_g + N_f \epsilon_q
= - 0.1273 \bar \mu^4 + N_f {3 \over 8 \pi^2} \bar \mu^4 \times I_q(x_0,0),    
\end{equation}
where $I_q(x_0,0)$ is given in (3.4) and $N_f$ is a number of light flavors. In
derivation of the YM part of the vacuum energy density, $\epsilon_g$, we use the relation  $p^2_0 = x_0^{min} \bar \mu^2$ (which once more shows the finiteness of $p_0$ since $\bar \mu$ is always finite) in order to express now the vacuum energy density in terms of the fundamental scale
parameter $\bar \mu$ and the constant of integration $x_0$ which in general need not to be numerically the same as in quenched QCD ($x_0^{min}$). That is why 
the nonperturbative effective scales for quenched ($N_f=0$) and full QCD are different while the fundamental scale parameter of our approach $\bar \mu$ is unique and its numerical value is to be found from good physical observable in full QCD only. The corresponding total effective potential at fixed $p_0$,
$ \Omega = - 0.1273 x_0^{-2} + N_f \Omega_q$, has no 
local minimum. Thus the injection of quark degrees of freedom lifts the nonperturbative quenched QCD vacuum from its stationarity.                            
How to choose the scale-setting scheme in order to numerically determine it is 
the subject for the next section.

\section{ The scale-setting scheme}

 The main problem now is to set the scale at which our calculations should 
be done. In our approach it means the choice of a reasonable value for the nonperturbative scale parameter $\bar \mu$ in (3.13). Only the requirement is that a scale-setting scheme should be physically well-motiviated since
$\bar \mu$  determines the physical scale of the nonperturbative dynamics in our approach, and it should not be arbitrarily large.                            
In our previous publications [19,26,39] the expressions for basic chiral QCD parameters such as pion decay constant, $F$ ($\equiv F^o_{\pi}$), the dynamically
generated quark mass $m_d$ (defined as the inverse of the quark propagator at zero momentum) and quark condensate ${\langle \overline qq \rangle}_0$ were derived:

\begin{equation}
F^2 = {3\over {8 \pi^2}} \bar \mu^2 \times I (x_0, 0) = 
           {3\over {8 \pi^2}} \bar \mu^2 \times
             \int^{x_0}_0 dx \,{ xB^2(x_0, x) \over
             {\{xA^2(x) + B^2(x_0, x)\}}},
\end{equation}

\begin{equation}
m_d = \bar \mu \bigl\{ B^2(x_0,0)\bigr\}^{-1/2},
\end{equation}

\begin{equation}
{\langle \overline qq \rangle}_0 = - {3\over {4\pi^2}} \bar \mu^3 \times I_{con}(x_0, 0) = - 
{3\over {4\pi^2}} \bar \mu^3  \times {\int}^{x_0}_0 dx\,{xB(x_0, x)}.
\end{equation} 
Here, as in (3.3-3.4), we need to identify the UV cut-offs with the constant of
integration $x_0$. In order to determine the fundamental mass scale parameter $\bar \mu$, 
which characterizes the region where confinement, DCSB and
other nonperturbative effects are dominant, we propose
to use the following bounds for the pion decay constant in the chiral limit
 
\begin{equation}
87.2 \leq  F^o_{\pi} \leq 93.3  \ (MeV).
\end{equation} 
The pion decay constant in the chiral limit, evidently, can not exeed its experimental value, so the upper bound in (4.4) is uniquely well-fixed. The lower
bound in (4.4) is fixed from the chiral value of the pion decay constant as obtained in Ref. [42], namely $F^o_{\pi} = (88.3 \pm 1.1) \ MeV$, which, obviously
, satisfies
(4.4). The value $F_{\pi} = 92.42 \ MeV$, advocated in Ref. [43], also satisfies these bounds. We think that chosen interval covers all the realistic values
of the chiral pion decay constant. In any case it is always possible to change 
lower bound to cover any requested value of the pion decay constant in the chiral limit. This bound is chosen as unique input data in our numerical investigation of chiral QCD.
The pion decay constant is a good experimental number since it is a directly
measurable quantity in contrast, for example, to the quark condensate or dynamically generated quark mass.
For this reason our choice (4.4) as input data opens up
the possibility of reliably estimating the deviation of the chiral
values from their "experimental" (empirical) phenomenologically determined
values of various physical quantities which can not be directly measured.      
Thus to assign definite values to the physical
quantities in the chiral limit is a rather delicate question (that is why we prefer to use and obtain bounds for them rather than the definite values). At
the same time it is a very important theoretical
limit which determines the dynamical structure of low-energy QCD.

What is necessary now is only to determine the constant of integration $x_0$.
For this aim, let us remind [19,25,26] that DCSB at the fundamental quark level
can be implemented by the following commutation relation

\begin{equation}
\{ S^{-1}(p), \gamma_5 \}_+ = i \gamma_5 2 \overline B(-p^2) \ne 0,
\end{equation} 
so that the $\gamma_5$ invariance of the quark propagator is broken and the
measure of this breakdown is the double of the dynamically generated quark mass
function, $2 \overline B(-p^2)$. Let us underline that this relation does not 
depend on gauge and renormalization point choices, by definition, i. e . in any
gauge and for any renormalization point this relation holds. In particular, denoting as usual the dynamically generated quark mass as the inverse of the quark
propagator (quark self-energy) at zero point, this quantity, $2 \overline B(0)$, can be defined as a scale of DCSB at the fundamental quark level, namely

\begin{equation}
\Lambda_{CSBq} =  2 \overline B(0) = 2 m_d.
\end{equation}
It is worthwhile to emphasize that $\Lambda_{CSBq}$ and $m_d$ have direct physical sense within our solutions to the quark SD equation since our approach to the deep IR (confinement) region is gauge invariant and free of ghost complications (that was discussed above in section 2). Thus we analyse our numerical data
at a scale where DCSB occurs at the fundamental quark level (remember also the 
discussion in the main body of the text after Eqs. (2.3-2.4)). What is needed in this case is only to simply identify our fundamental mass scale parameter
$\bar \mu$ with $\Lambda_{CSBq}$, i. e. to put $\bar \mu \equiv \Lambda_{CSBq} = 2 m_d$. From Eq. (4.2) then it immediately follows that the constant of integration $x_0$ is equal to

\begin{equation}
x_0 = 1.53
\end{equation}
and it is different from the quenched QCD value $x_0^{min}$ (3.11) at which stationary state occurs there. Without the existence of the stationary state (localization of the minimum) one has no criterion how to numerically distinguish them. The advantage of this scale-setting scheme is that,
on one hand, it is based on exact relation (4.6), on the other hand, numerical 
values of the mass scale parameter $\bar \mu$ are extracted from the pion decay
constant which, as it was emphasized above, is a good experimental quantity. Indeed, substituting (4.7) into Eq. (4.1) and combining it with (4.4), one immediately obtains values of the fumdamental mass scale parameter  $\bar \mu$.  
This makes it possible to numerically calculate all physical quantities considered in our work. These results are presented in Table 1 exept for the vacuum 
energy density.   

It is interesting to note that the information on the dimensionless constant of
integration $x_0$ is taken from (4.2) and (4.6) on account of the identification $\bar \mu \equiv \Lambda_{CSBq} = 2 m_d$, i. e., in fact, inevitable transformation of the pair of independent parameters $x_0, \ \bar \mu$ into the pair of
$\Lambda_{CSBq} (m_d), \ \bar \mu$ takes place within our scale-setting calculation scheme. This is also a manifestation of the phenomenon of the "dimensional
transmutation" [44], which occurs whenever a massless theory aquires masses dynamically. It is a general feature of spontaneous symmetry breaking in field theories.

\section{Numerical results and conclusions}

That the QCD ground state (vacuum) has a stationary state for pure gluodynamics
(quenched QCD) is a direct consequence of the existence and importance of the nonperturbative $q^{-4}$-type quantum, topologically nontrivial dynamical excitations of the gluon field configurations. Moreover, the inclusion of quark degrees of freedom lifts the nonperturbative YM vacuum from its stationarity. Our values for the chiral QCD vacuum energy density are:

\begin{eqnarray}
\epsilon &=& - ( 0.01425 - 0.00196 N_f) \ GeV^4, \\
\epsilon &=& - ( 0.01087 - 0.00150 N_f) \ GeV^4, 
\end{eqnarray}
where the first and second values are due to the upper and the lower bounds
in (4.4), respectively. As was mentioned in the Introduction, there exists also
a contribution to the vacuum energy density at the classical level given by the
instanton-type nonperturbative fluctuations of gluon fields. Within RILM [5] for dilute ensemble it is:                        

\begin{equation}
\epsilon_I = - {b \over 4}n = - {b \over 4} \times 1.0 \ fm^{-4} = - ( 0.00417 - 0.00025 N_f) \ GeV^4, 
\end{equation}
where $b = 11 - (2/3)N_f$ is the first coefficient of the $\beta$-function and $n$ is the density of the instanton-type fluctuations in the QCD vacuum. This expression was postulated via the trace anomaly relation using the weak coupling
limit solution to the above mentioned $\beta$-function as well as the phenomenological value of the gluon condensate [45]. Recently, in quenched 
($N_f=0$) lattice QCD by using the so-called "cooling" method the role of instantons in the QCD vacuum was investigated [46]. In particular, it was found that
the instanton density should be $n = (1 + \delta)fm^{-4}$, where $\delta \simeq
(0.3 - 0.6)$, depending on cooling steps. 
We find that our values for the vacuum energy density (5.1-5.2) are an order of
magnitude bigger than the instanton-based RILM (in various modifications) [5,46] can provide at all, Eq. (5.3). This can be
easly understood indeed since our approach is relevant in the strong coupling 
limit while instantons are weak coupling limit phenomena.   
                   
Stationarity of the nonperturbative YM vacuum and the numerical results for the
chiral QCD vacuum energy density (5.1-5.2) which for the first time have been calculated from first principles (not postulated via the trace anomaly relation 
as (5.3)) are our main results.\footnote{We use the 
terms quenched QCD, YM theory and pure gluodynamics as synonyms, as well as the
ground state and the nonperturbative vacuum. However the ground state of the nonperturbative YM vacuum means its stationary state within our terminology. The 
existence of the stationary state in the nonchiral (real) QCD vacuum due to 
$q^{-4}$-type excitations of gluon fields remains an open question yet.}
It is also clearly shown in our work that precisely the nonperturbative $q^{-4}$-type quantum, topologically nontrivial dynamical excitaions of the gluon field configurations in the QCD vacuum is mainly responsible for quark confinement 
and other nonperturbative phenomena (at least at the fundamental quark level). 
This is our answer to the question arised by DeGrand et al. in Ref. [12] and by
Negele in Ref. [46].  
At the same time, it is very interesting to directly identify this type of field configurations by lattice simulations. 

 Recently the effects of nonperturbative QCD in the nucleon structure functions
were investigated. A universal mass scale parameter $m_a \simeq 470 \ MeV$ of the nonperturbative dynamics in QCD was obtained [47]. This is in rather good agreement with our numerical results for the physical mass scale parameter $\bar \mu$ (see Table 1) taking into account completely different physical observables have been analysed.
The chiral symmetry breaking scale within exact renormalization group approach 
[48] is found to be around $(400-500) \ MeV$. Again rather close to our numerical results for $\bar \mu$.                                                  

Recent quenched lattice result for the chiral condensate which does not depend 
neither on the scale nor on the renormalization scheme [49] (in our notations),
${\langle \overline qq \rangle}_0 = - (206 \ MeV )^3$, nicely satisfies our bounds (see Table 1). For simplicity, here and below  we  
show only central values, omitting errors due to statistics, renormalization and lattice calibration [49]. Recalculated at the conventional scale $ 2 \ GeV$  
(with the help of expressions given in [49] for $\overline {MS}$ scheme) it becomes ${\langle \overline qq \rangle}_0 ( 2 \ GeV) = - (245 \ MeV )^3$. This is 
in fair agreement with recent phenomenological determination from QCD sum rules
[50], ${\langle \overline qq \rangle}_0 ( 1 \ GeV) = - (229 \ MeV )^3$, which 
at the scale $ 2 \ GeV$ becomes ${\langle \overline qq \rangle}_0 ( 2 \ GeV) = - (242 \ MeV )^3$. Thus our bounds, formally recalculated at the scale $2\ GeV$, $ (230 \ MeV)^3 \leq -  {\langle \overline qq \rangle}_0 ( 2 \ GeV) \leq     
(245 \ MeV)^3$, being in fair agreement with the above discussed phenomenological and lattice data, favor a large quark condensate advocated by the chiral perturbation theory (see discussion in [51]).
%The dynamically generated quark masses are about three hundred $MeV$, which ar%e quite resonable while the constituent quark masses are usually estimated as %about four hundred $MeV$. Solving the SD equation, describing the propagation %of a single quark, we can not account for the constituent mass, only for
%the effective one which becomes dynamical in the chiral limit. Constituent mas%s is the subject of the bound-state problem.  

 In Table 1, the nonperturbative effective scale for quenched QCD, defined as  
$p_0^2 = x_0^{min} \bar\mu^2$, is denoted as $\Lambda_{eff}^{(0)}$ while for the full QCD, defined as $p_0^2 = x_0 \bar\mu^2$ (and numerical value of $x_0$ is
given in (4.7)), is denoted as $\Lambda_{eff}^{(N_f)}$. As a subject for brief 
discussion, let us note that an alternative interpretation of these scales is
also possible. It is obvious that $\bar\mu$ can be identified with the scale
at which chiral symmetry is broken dynamically at the fundamental quark level, i. e. $\bar\mu = \Lambda_{CSBq}$. This was the heart of our setting-scale calculation scheme. It makes sense in addition to identify our effective scales with
confinement scales, i. e. to put  $\Lambda_{eff}^{(0)} = \Lambda_C^{(0)} \equiv
\Lambda^C_{YM}$ and $\Lambda_{eff}^{(N_f)} = \Lambda_C^{(N_f)} \equiv \Lambda^C_{QCD}$ for quenched (YM) and full QCD, 
respectively. If QCD confines color, then these two scales in principle need not to be the same, while $\bar\mu = \Lambda_{CSBq}$ remains unique as coming from full QCD. Thus within this interpretation there is a hierarchy of scales at the fundamental quark level, namely  $\Lambda^C_{YM} > \Lambda^C_{QCD} > \Lambda_{CSBq}$. The ratio between them are given by the corresponding constants of integration. At the same time all these scales are less than the chiral symmetry
breaking scale at the hadronic level, $\Lambda_{{\chi}{SB}} \leq 4 \pi F_{\pi} \simeq 1.2 \ GeV$ [52] (see also Ref. [53] and references therein). At this stage, the relation of our scales to the QCD scale parameter, $\Lambda_{QCD}$, remains, of course, unanswered. We hope to shed some light on this possible relation taking into account UV terms in the quark SD equation (2.3-2.4) where  
$\Lambda_{QCD}$ naturally appears via the perturbative renormalization group logarithms.

For the sake of completeness, let us present the numerical values 
of basic integrals (containing explicitly quark degrees of freedom) which allows one to check our numerical results. These integrals are: $I_q(1.53, 0) = 0.46$, which is given in (3.4), $I (1.53, 0) = 0.6876$, which is given in (4.1) and
$I_{con} (1.53, 0) = 0.6837$, which is given in (4.3). Evidently, these  
integrals can be calculated to any requested accuracy. And finally, the numerical values of the bag constant $B$, defined as the difference between the perturbative and nonperturbative vacua [54], are given now by the relation $B = -  \epsilon$ (since the perturbative vacuum is normalized to zero) and can be explicitly evaluated from (5.1),  

\begin{eqnarray}
B (N_f=0) = 0.01425 \ GeV^4 = (345.5 \ MeV)^4 = 1.85 \ GeV/fm^3,\\
B (N_f=1) = 0.0123 \ GeV^4 = (333 \ MeV)^4 = 1.60 \ GeV/fm^3,\\
B (N_f=2) = 0.0103 \ GeV^4 = (318.6 \ MeV)^4 = 1.34 \ GeV/fm^3,\\
B (N_f=3) = 0.00838 \ GeV^4 = (302.5 \ MeV)^4 = 1.09  \ GeV/fm^3,
\end{eqnarray}
and from (5.2),

\begin{eqnarray}
B (N_f=0) = 0.01087 \ GeV^4 = (322.9 \ MeV)^4 = 1.40 \ GeV/fm^3,\\
B (N_f=1) = 0.00937 \ GeV^4 = (311.1 \ MeV)^4 = 1.22 \ GeV/fm^3,\\
B (N_f=2) = 0.00787 \ GeV^4 = (297.8 \ MeV)^4 = 1.02 \ GeV/fm^3,\\
B (N_f=3) = 0.00637 \ GeV^4 = (285.5 \ MeV)^4 = 0.83  \ GeV/fm^3.
\end{eqnarray}

 The numerical investigation of chiral QCD topology within our model and its comparison with instanton based models is the subject for subsequent paper.

\acknowledgements

One of the authors (V.G.) is grateful to the late Prof. V.N. Gribov for many
useful remarks and discussions on nonperturbative QCD. This work is dedicated to his memory. It was supported by Special COE grant of the Ministry of Education, Sience and Culture of Japan.

%\vfill

%\eject

%\vfill

%\eject

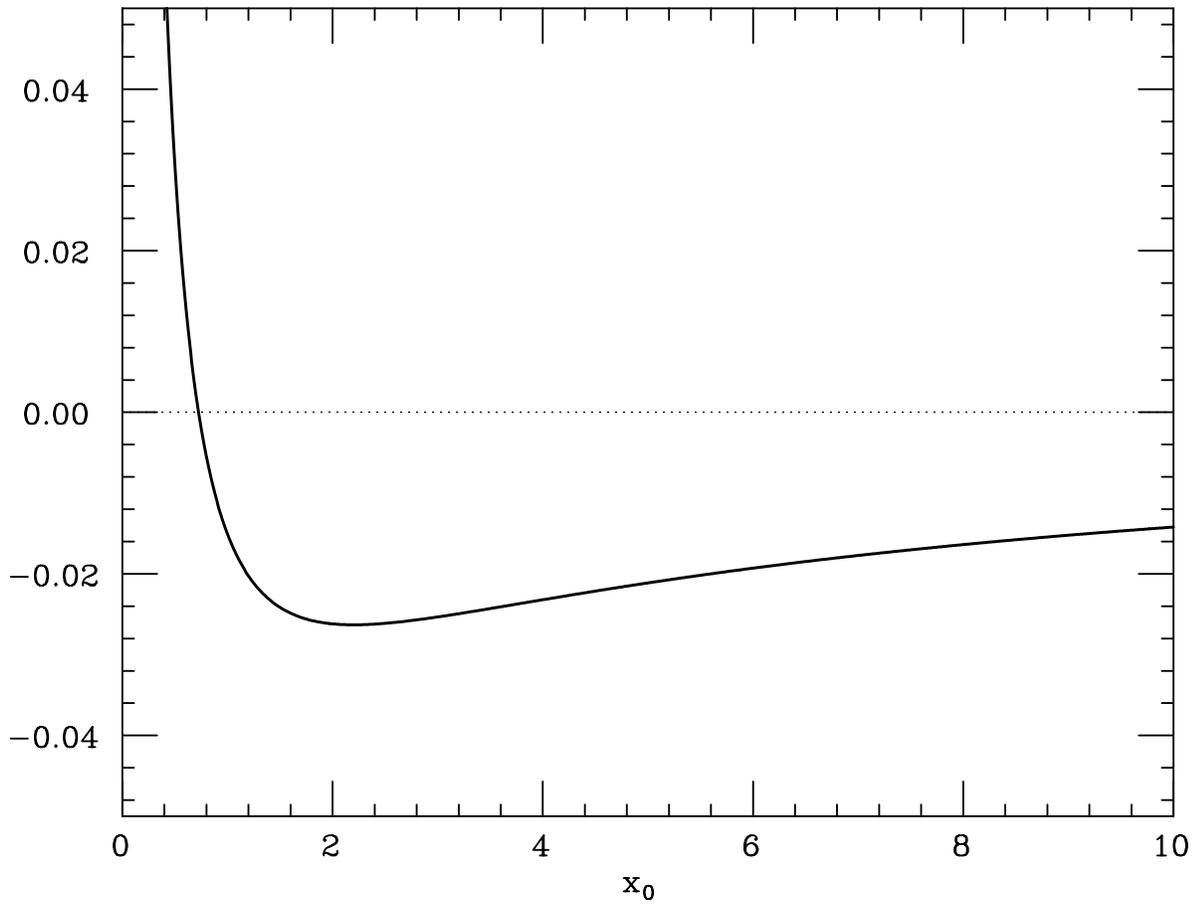
\begin{figure}
\caption{The effective potential $\Omega_g$ in (3.10) as a function of $x_0$.  
         A local minimum appears at $x_0^{min} = 2.20$ and zero at $x_0^{0} = 0.725$. $\Omega_g$ falls as $\sim - x_0^{-1}$ asymptotically at large $x_0$. }  
\end{figure}


\begin{references}
\bibitem{1}
   P.Olesen, Phys.Scr., {\bf 23} (1981) 1000
\bibitem{2}
   L.S.Celenza, C.-R. Ji and C.M.Shakin, Phys.Rev., {\bf D36} (1987) 895
\bibitem{3}
   H.G.Dosch and Yu.A.Siminov, Phys.Lett., {\bf B205} (1988) 339
%  Yu.A.Siminov, Nucl.Phys., {\bf B307} (1988) 512
\bibitem{4}
   J.Hosek and G.Ripka, Z.Phys., {\bf A354} (1996) 177
\bibitem{5}
   T.Schafer and E.V.Shuryak, Rev.Mod.Phys., {\bf 70} (1998) 323
\bibitem{6}
   G. 't Hooft, Phys.Rev., {\bf D14} (1976) 3432
\bibitem{7}
   C.G.Callan, Jr, R.F.Dashen and D.J.Gross, Phys.Rev., {\bf D17} (1978) 2717;    \\
   D.I.Dyakonov and V.Yu.Petrov, Nucl.Phys., {\bf B272} (1986) 457
\bibitem{8}
   A.A.Belavin, A.Polyakov, A.Schwartz and Y.Tyupkin , Phys.Lett., {\bf B59} (1   975) 85
\bibitem{9}
   Confinement, Duality, and Nonperturbative Aspects of QCD (ed. by P.van Baal)
   NATO ASI Series B: Physics, vol. 368; Non-Perturbative QCD, Structure of the
   QCD vacuum, (ed. by K-I.Aoki, O.Miymura and T.Suzuki), Prog.Theor.Phys.Suppl., {\bf 131} (1998) 
\bibitem{10}
   V.N.Gribov, LUND preprint, LU TP 91-7, (1991) 1-51   
\bibitem{11}
   R.J.Crewther, Nucl.Phys., {\bf B209} (1982) 413;   \\
   R.Jackiw and C.Rebbi, Phys.Rev.Lett., {\bf 37} (1976) 172
\bibitem{12}  
   T.DeGrand, A.Hasenfratz and T.G.Kovacs, Nucl.Phys., {\bf B505} (1997) 417;     \\
   D.A.Smith and M.J.Teper, Phys.Rev., {\bf D58} (1998) 014505
\bibitem{13}
   J.L.Gervais and A.Neveu, Phys.Rep., {\bf C23} (1976) 240; \\
   L.Susskind and J.Kogut, Phys.Rep., {\bf C23} (1976) 348; \\
   T.Banks and A.Casher, Nucl.Phys., {\bf B169} (1980) 103
\bibitem{14}
   D.J.Gross and F.Wilczek, Phys., {\bf D8} (1973) 3633; \\
   S.Weinberg, Phys.Rev.Lett., {\bf 31} (1973) 494; \\
   H.Georgi and S.Glashow, Phys.Rev.Lett., {\bf 32} (1974) 438
\bibitem{15}
   S.Mandelstam, Phys.Rev., {\bf D20} (1979) 3223
\bibitem{16}
   M.Baker, J.S.Ball and F.Zachariasen, Nucl.Phys., {\bf B186} (1981) 531, 560;
   ibid, {\bf B226} (1983) 455  
\bibitem{17}
   N.Brown and M.R.Pennington, Phys.Rev., {\bf D39} (1989) 2723; \\
   B.A.Arbuzov, Sov.Jour.Part.Nucl., {\bf 19} (1988) 1; \\
   D.Atkinson, H.Boelens, S.J.Hiemastra, P.W.Johnson, 
   W.E.Schoenmaker and K.Stam, J.Math.Phys., {\bf 25} (1984) 2095; \\
   C.D.Roberts and A.G.Williams, Prog.Part.Nucl.Phys., {\bf 33} (1994) 477
\bibitem{18}
   H.Pagels, Phys.Rev., {\bf D15} (1977) 2991
\bibitem{19}
   V.Gogohia, Gy.Kluge and M.Prisznyak, hep-ph/9509427; \\
   V.Gogohia and H.Toki, in preparation
\bibitem{20}
   V.P.Nair and C.Rosenzweig, Phys.Rev., {\bf D31} (1985) 401
\bibitem{21}
   H.Narnhofer and W.Thirring, in Springer Tracks Mod.Phys., {\bf 119} (1990) 1
\bibitem{22}
   K.Johhson, L.Lellouch and J.Polonyi, Nucl.Phys., {\bf B367} (1991) 675 
\bibitem{23}
   K.B.Wilson, Phys.Rev., {\bf D10} (1974) 2445; \\ 
   M.Bander, Phys.Rep., {\bf 75} (1981) 205;
   G.B.West, Phys.Lett., {\bf B115} (1982) 468
\bibitem{24}
   U.Ellwanger, M.Hirsch and A.Weber, Eur.Phys.J., {\bf C1} (1998) 563
\bibitem{25}
   V.Sh.Gogohia, Phys.Rev., {\bf D40} (1989) 4157; ibid, {\bf D41} (1990) 3279;   \\
   V.Sh.Gogohia, Inter.Jour.Mod.Phys., {\bf A9} (1994) 759
\bibitem{26}
   V.Gogohia, Gy.Kluge and M.Prisznyak,  Phys.Lett., {\bf B368} (1996) 221
\bibitem{27}
   W.Marciano and H.Pagels, Phys.Rep., {\bf C36} (1978) 137
\bibitem{28}
   F.Strocchi, Phys.Lett., {\bf B62} (1976) 60
\bibitem{29}
   A.I.Alekseev and B.A.Arbuzov, Phys.Atom.Nucl., {\bf 61} (1998) 264
\bibitem{30}
   G.'t Hooft and M.Veltman, Nucl.Phys., {\bf B44} (1972) 189; \\
   C.G.Bollini and J.J.Giambiagi, Nuovo Cim., {\bf B12} (1972) 20
\bibitem{31}
   J.N.Gelfand and G.E.Shilov, Generalized Functions,
   (Academic Press, NY, 1964)
\bibitem{32}
   G.Preparata, Phys.Rev., {\bf D7} (1973) 2973
\bibitem{33}
   G.'t Hooft, Nucl.Phys., {\bf B75} (1974) 461
\bibitem{34}
   K.D.Born, E.Laerman, R.Sommer, T.F.Walsh and P.M.Zerwas, Phys.Lett., {\bf B3   29} (1994) 325; \\
   V.M.Miller, K.M.Bitar, R.C.Edwards and A.D.Kennedy, Phys.Lett., {\bf B335} (   1994) 71
\bibitem{35}
   J.M.Cornwall, Phys.Rev., {\bf D22} (1980) 1452
\bibitem{36}
   J.M.Cornwall, R.Jackiw and E.Tomboulis, Phys.Rev., {\bf D10} (1974) 2428
\bibitem{37}
   A.Barducci, R.Casalbuoni, S.De Curtis, D.Dominici and R.Gatto,
   Phys.Rev. {\bf D38} (1988) 238
\bibitem{38}
   N.K.Nielsen and P.Olesen, Nucl.Phys., {\bf B144} (1978) 376
\bibitem{39}
   V.Sh.Gogohia, Gy.Kluge and B.A.Magradze, Phys.Lett., {\bf B244} (1990) 68
\bibitem{40}
   M.E.Peskin and D.V.Schroeder, An Introduction to Quantum Field Theory       
   (Addison-Wesley, Advanced Book Program, 1995)
\bibitem{41}
   V.Gogohia, hep-ph/9806251 
\bibitem{42}
   P.Gerber and H.Leutwyler, Nucl.Phys., {\bf B321} (1989) 387
\bibitem{43}
   B.R.Holstein, Phys.Lett., {\bf B244} (1990) 83; \\
   W.J.Marciano and A.Sirlin, Phys.Rev.Lett., {\bf 71} (1993) 3629
\bibitem{44}
   S.Colemen and E.Weinberg, Phys.Rev., {\bf D7} (1973) 1888; \\
   D.J.Gross and A.Neveu, Phys.Rev., {\bf D10} (1974) 3235
\bibitem{45}
   M.A.Shifman, A.I.Vainstein and V.I.Zakharov, Nucl.Phys., {\bf B147} (1979) 385, 448
\bibitem{46}
   M.-C.Chu, J.M.Grandy, S.Huang and J.W.Negele, Phys.Rev., {\bf D49} (1994)   
   6039; \\ 
   J.W.Negele, hep-lat/9810053   
\bibitem{47}
   Y.Mizuno, Nucl.Phys., {\bf A629} (1998) 55c
\bibitem{48}
   D.-U.Jungnickel and C.Wetterich, hep-ph/9710397
\bibitem{49}
   L.Guisti, F.Rapuano, M.Taveli and A.Vladikas, hep-lat/9807014
\bibitem{50}
   H.G.Dosch and S.Narison, Phys.Lett., {\bf B417} (1998) 173
\bibitem{51}
   G.Ecker, hep-ph/9805500
\bibitem{52}
   A.Manohar and H.Georgi, Nucl.Phys., {\bf B234} (1984) 189; \\
   J.F.Donoghue, E.Golovich and B.R.Holstein, Phys.Rev., {\bf D30} (1984) 587
\bibitem{53}
   V.Sh.Gogohia, Inter.Jour.Mod.Phys., {\bf A9} (1994) 605
\bibitem{54}
   M.Chanowitz and S.Sharpe, Nucl.Phys., {\bf B222} (1983) 211
\end{references}
\end{document}